\documentclass[journal]{IEEEtran}

\usepackage{cite}
\usepackage{amsmath,amssymb,amsfonts}
\usepackage{graphicx}
\usepackage{booktabs}
\usepackage{multirow}
\usepackage{microtype}

\usepackage{algorithm}
\usepackage{algorithmic}

\usepackage{tikz}
\usetikzlibrary{positioning,arrows.meta}
\usepackage{stfloats}

\begin{document}

\title{Impact of Validation Strategy on Machine Learning Performance in EEG-Based Alcoholism Classification}

\author{
Tahir~Cetin~Akinci,~\IEEEmembership{Senior Member,~IEEE}, 
Yuksel~Celik,~\IEEEmembership{Senior Member,~IEEE}, 
and Omer~Faruk~Ertugrul%,~\IEEEmembership{Senior Member,~IEEE}
\thanks{
T.~C.~Akinci is with the Department of Electrical and Computer Engineering, Florida Polytechnic University, Lakeland, FL 33805, USA (e-mail: takinci@floridapoly.edu).

Y.~Celik is with the Department of Electrical and Computer Engineering, University at Albany, State University of New York, Albany, NY 12222, USA (e-mail: ycelik@albany.edu).

O.~F.~Ertugrul is with the Department of Electrical and Electronics Engineering, Batman University, Batman 72060, Turkiye (e-mail: omerfaruk.ertugrul@batman.edu.tr).
}
}

\maketitle

%%%%%%%%%%%%%%%%%%%%%%%%%%%%%%%%%%%%%%%%%%%%%%%%%%%%%%%%%
%%%%%%%%%%%%%%%%%%%%%%%%%%%%%%%%%%%%%%%%%%%%%%%%%%%%%%%%%
\begin{abstract}
Electroencephalography (EEG) provides a non-invasive and cost-effective approach for analyzing neural patterns associated with alcohol dependence. However, reported classification performance in EEG-based alcoholism studies varies considerably, often due to differences in validation strategies rather than intrinsic model capability. This study presents a validation-aware machine learning framework to systematically assess the impact of evaluation methodology on classification performance. A balanced multi-channel EEG dataset consisting of 300 trials (150 alcoholic, 150 control) was analyzed using a structured feature representation that integrates statistical descriptors and spectral band interactions. Five classifiers, including support vector machines (linear and radial basis function kernels), random forest, k-nearest neighbors, and AdaBoost, were evaluated under both standard cross-validation and nested cross-validation protocols. The results demonstrate that conventional validation with global hyperparameter tuning introduces optimistic bias. In particular, SVM with radial basis function kernel exhibited a performance decrease of approximately 5\% under nested cross-validation, indicating overestimation in standard evaluation. In contrast, ensemble-based methods showed more stable generalization, with AdaBoost achieving the highest performance, reaching 78.3\% accuracy ($\pm$4.25), an AUC of 0.868, and balanced sensitivity (78.67\%) and specificity (81.33\%). These findings highlight that the validation strategy is a primary determinant of perceived model performance in high-dimensional EEG classification. Statistical analysis using McNemar's test further shows that most performance differences between models are not statistically significant, reinforcing the importance of careful interpretation of classification results. The proposed framework provides a reproducible and methodologically robust basis for evaluating machine learning models in biomedical signal analysis.
\end{abstract}

%%%%%%%%%%%%%%%%%%%%%%%%%%%%%%%%%%%%%%%%%%%%
\begin{IEEEkeywords}
Electroencephalography (EEG), Alcoholism Detection, Nested Cross-Validation, Data Leakage, Ensemble Learning, AdaBoost, Support Vector Machine, Feature Engineering, Validation Bias, Biomedical Signal Processing
\end{IEEEkeywords}
%%%%%%%%%%%%%%%%%%%%%%%%%%%%%%%%%%%%%%%%%%%%

%%%%%%%%%%%%%%%%%%%%%%%%%%%%%%%%%%%%%%%%%%%
\section{Introduction}
%%%%%%%%%%%%%%%%%%%%%%%%%%%%%%%%%%%%%%%%%%%

Alcohol use disorder (AUD) represents a major global health problem, affecting millions of individuals worldwide and leading to significant cognitive, behavioral, and neurological impairments \cite{rangaswamy2004,porjesz2005,coutin2006}. Early and reliable detection of alcohol-related neural alterations is essential for effective diagnosis and intervention, motivating the search for quantitative EEG-based biomarkers \cite{acharya2014,zhang2023review}. Prior neurophysiological studies have shown that alcohol dependence alters oscillatory brain activity, particularly within the beta band, providing a measurable basis for objective assessment \cite{rangaswamy2004,porjesz2005}.

Electroencephalogram signals provide a non-invasive and cost-effective means of monitoring brain activity and have been extensively used in the analysis of neurological disorders and substance-related conditions \cite{acharya2014,mumtaz2017,AkinciALS2018}. However, EEG signals are inherently nonlinear, nonstationary, and stochastic, making their analysis challenging. Classical signal processing techniques, including spectral analysis and time--frequency representations, have been widely employed to capture oscillatory patterns and transient behaviors associated with cognitive processes \cite{AkinciWelding2010,AkinciCable2009,AkinciCeramic2011}. In particular, time--frequency methods enable localized analysis of signal energy distributions, offering improved insight into dynamic neural processes compared to purely global spectral approaches \cite{AkinciWelding2010,AkinciCeramic2011,Li2022Transformer}.

Such signal processing methodologies have demonstrated strong capability in various engineering applications involving nonstationary signals, including fault detection in electrical systems and transient signal characterization \cite{AkinciCable2009,AkinciWelding2010,AkinciCeramic2011}. These approaches emphasize the importance of analyzing localized energy distributions and structural variations in complex signals, which is also critical for understanding EEG dynamics.

In parallel, recent advances have introduced data-driven approaches in EEG analysis, where machine learning (ML) and artificial neural networks (ANNs) are used to automatically extract discriminative features and perform classification tasks \cite{acharya2014,mumtaz2017,AkinciALS2018}. Applications such as epileptic seizure detection, neurological disorder assessment, and gait analysis demonstrate the effectiveness of ML-based EEG analysis in handling high-dimensional and complex data \cite{acharya2014,AkinciALS2018}. More recent studies have shown that deep learning architectures, attention mechanisms, and domain adaptation techniques significantly improve EEG classification robustness across subjects and recording conditions \cite{zhang2023review,roy2023attention,wang2024domain,Gupta2023Interpretable,Wang2023MultiDomain}. Among ensemble methods, AdaBoost \cite{freund1997} has demonstrated strong performance in biomedical classification tasks by adaptively focusing on hard-to-classify samples through iterative reweighting, making it particularly suitable for imbalanced or noisy data. Random Forest \cite{breiman2001} employs bootstrap aggregation (bagging) of decision trees, where each tree is trained on a random subset of features and data samples. This ensemble approach reduces variance and improves stability compared to individual decision trees, making it robust to overfitting in high-dimensional feature spaces. These findings highlight the growing importance of advanced ML models in biomedical signal processing.

Despite the growing interest in EEG-based alcoholism detection, a consistent limitation across existing studies is the lack of methodological standardization in both feature representation and evaluation strategy. Many approaches rely on narrowly defined feature sets or model-specific optimizations that may not generalize well across subjects or recording conditions. In addition, inconsistencies in validation protocols hinder fair comparison between studies and may obscure the true effectiveness of proposed models. These issues limit the development of clinically reliable EEG biomarkers and reduce the reproducibility of reported results. Therefore, there is a clear need for frameworks that combine interpretable feature engineering with rigorous and transparent evaluation methodologies.

In particular, many existing studies focus primarily on maximizing classification accuracy, often overlooking the impact of validation strategy on reported results. Improper evaluation procedures, such as global hyperparameter tuning combined with standard cross-validation, may introduce data leakage and lead to overly optimistic performance estimates \cite{varma2006,cawley2010}. This issue is especially pronounced in high-dimensional EEG feature spaces, where model behavior is highly sensitive to hyperparameter configurations, making rigorous validation essential for reliable performance assessment \cite{Roy2023SelfSupervised,Zhang2023GCN,varma2006}. The EEG signals used in this study were obtained from the publicly available UCI Machine Learning Repository, originally collected by Begleiter at the Neurodynamics Laboratory of SUNY Health Center Brooklyn \cite{uci_eeg}.

Motivated by these observations, this study proposes a validation-aware machine learning framework for EEG-based classification of alcoholism. The proposed approach integrates statistical descriptors and spectral-band-based features into a structured multi-channel representation, while explicitly addressing evaluation bias through nested cross-validation \cite{Zhang2023GCN,cawley2010}. Unlike purely data-driven approaches, the framework emphasizes physically interpretable feature construction and reproducible evaluation protocols, ensuring that reported performance reflects true generalization capability.

%\vspace{-6cm}

%%%%%%%%%%%%%%%%%%%%%%%%%%%%%%%%%%%%%%%%%%%%%%%%%%%%%%%%%%%%%

%%%%%%%%%%%%%%%%%% SECTION 2 %%%%%%%%%%%%%%%%%%%%%%%%%%%%%%
\section{Proposed Validation-Aware EEG Analysis Framework}
%%%%%%%%%%%%%%%%%%%%%%%%%%%%%%%%%%%%%%%%%%%%%%%%%%%%%%%%%%%
%\vspace{-3.55cm}

This study proposes a structured and validation-aware framework for EEG-based alcoholism classification, integrating signal processing, feature engineering, and machine learning within a unified pipeline \cite{rodrigues2019eeg,mukhtar2021cnn}. Unlike conventional approaches that primarily focus on classification accuracy, the proposed framework emphasizes physically interpretable feature representation and unbiased model evaluation \cite{sangle2025xai,ahmed2026stacked}. In addition, the framework is designed to ensure reproducibility by explicitly defining each processing stage and its role within the overall pipeline \cite{ahmed2026stacked,patil2025eeg}.

\begin{figure}[H]
\centering
\includegraphics[width=0.55\columnwidth]{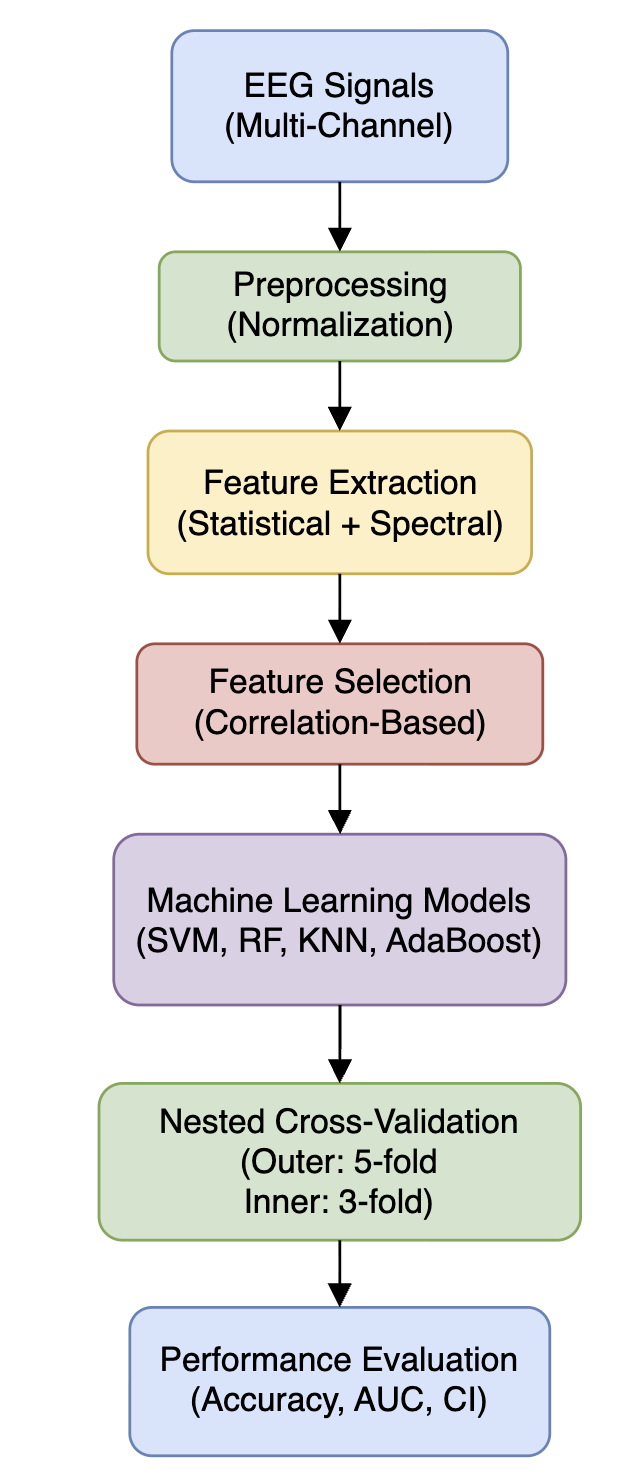}
\caption{Proposed validation-aware EEG analysis framework. The pipeline consists of signal preprocessing, multi-channel feature extraction, feature selection, and classification using multiple machine learning models. A nested cross-validation strategy is employed to ensure unbiased performance evaluation and prevent data leakage during hyperparameter optimization.}
\label{fig:framework}
\end{figure}

The overall architecture of the proposed methodology is illustrated in Fig.~\ref{fig:framework}. The pipeline consists of four main stages: signal preprocessing, multi-channel feature extraction, feature construction and selection, and classification under a nested cross-validation strategy \cite{tekale2026awsan,vaniya2025depression}. Each stage is designed to address specific challenges associated with EEG data, including variability across subjects, high dimensionality, and susceptibility to overfitting \cite{chhabra2026emotion}. This structure ensures both interpretability and robustness in evaluating model performance.

%%%%%%%%%% FIGURE 1 %%%%%%%%%%%%%%%%
%%%%%%%%%%%%%%%%%%%%%%%%%%%%%%%%%%

%%%%%%%%%%%%%%%%%%%%%%%%%%%%%%%%%%
\subsection{Signal Preprocessing}
%%%%%%%%%%%%%%%%%%%%%%%%%%%%%%%%%%%

EEG signals are first normalized to eliminate amplitude variability across channels and subjects \cite{saini2026stress}. Prior to normalization, each EEG segment is treated independently to avoid information leakage across samples \cite{vaniya2025depression}. Given a raw EEG signal $x_c(t)$ from channel $c$, normalization is performed as

\begin{equation}
\tilde{x}_c(t) = \frac{x_c(t)}{\max |x_c(t)|},
\label{eq:norm}
\end{equation}

where $\tilde{x}_c(t)$ denotes the normalized signal. As defined in (\ref{eq:norm}), this operation ensures numerical stability and allows consistent comparison across different EEG recordings by removing scale differences \cite{liu2018cdss}. This normalization is applied channel-wise, preserving the relative temporal structure of the signal while standardizing amplitude variations.

%%%%%%%%%%%%%%%%%%%%%%%%%%%%%%%%%%%%%%%%%%%%%%
\subsection{Multi-Channel Feature Extraction}
%%%%%%%%%%%%%%%%%%%%%%%%%%%%%%%%%%%%%%%%%%%%%%

To capture both temporal and spectral characteristics of EEG signals, a combination of statistical and frequency-domain features is extracted from each channel \cite{saini2026stress}. Feature extraction is performed on fixed-length EEG segments, ensuring consistency across all samples and preventing bias in feature computation \cite{patil2025eeg}. This dual-domain representation enables the model to capture both amplitude-based variations and oscillatory behavior \cite{chhabra2026emotion}.

%%%%%%%%%%%%%%%%%%%%%%%%%%%%%%%%%%%%%%%%%%%%%%
\subsubsection{Statistical Features}
%%%%%%%%%%%%%%%%%%%%%%%%%%%%%%%%%%%%%%%%%%%%%%

The first set of features describes the statistical properties of the signal \cite{liu2018cdss}:

\begin{equation}
\mu_c = \frac{1}{N} \sum_{t=1}^{N} x_c(t),
\label{eq:mean}
\end{equation}

\begin{equation}
\sigma_c = \sqrt{\frac{1}{N} \sum_{t=1}^{N} (x_c(t) - \mu_c)^2},
\label{eq:std}
\end{equation}

\begin{equation}
\kappa_c = \frac{1}{N} \sum_{t=1}^{N} \left( \frac{x_c(t) - \mu_c}{\sigma_c} \right)^4.
\label{eq:kurt}
\end{equation}

Here, (\ref{eq:mean}) represents the mean value, (\ref{eq:std}) the standard deviation, and (\ref{eq:kurt}) the kurtosis of the signal. These descriptors capture signal distribution, variability, and peak characteristics, which are known to be affected by neurological conditions \cite{patil2025eeg}. In particular, increased variability and higher-order statistics may reflect irregular neural firing patterns associated with alcohol-induced brain alterations \cite{rodrigues2019eeg}.

%%%%%%%%%%%%%%%%%%%%%%%%%%%%%%%%%%%%%%%%%%%%%%
\subsubsection{Spectral Features}
%%%%%%%%%%%%%%%%%%%%%%%%%%%%%%%%%%%%%%%%%%%%%%

The frequency-domain representation is obtained using the Fourier transform \cite{bhuvaneshwari2026efd}:

\begin{equation}
X_c(f) = \mathcal{F}\{x_c(t)\},
\label{eq:fft}
\end{equation}

and the corresponding spectral energy is defined as

\begin{equation}
E_c(f) = |X_c(f)|^2.
\label{eq:energy}
\end{equation}

Based on (\ref{eq:energy}), band-specific energies are computed over standard EEG frequency ranges \cite{tekale2026awsan}:

\begin{equation}
E_c^{b} = \int_{f_b^{low}}^{f_b^{high}} E_c(f)\, df,
\label{eq:band}
\end{equation}

where $b \in \{\delta, \theta, \alpha, \beta\}$. These band energies represent physiologically meaningful oscillatory components of EEG signals, where different frequency bands are associated with distinct cognitive and neurological functions \cite{mukhtar2021cnn}.

To eliminate scale dependency, normalized band energies are defined as

\begin{equation}
\tilde{E}_c^{b} = \frac{E_c^{b}}{\sum_{b} E_c^{b}},
\label{eq:normenergy}
\end{equation}

As expressed in (\ref{eq:normenergy}), this normalization enables comparison across subjects by representing relative energy distribution rather than absolute values \cite{saini2026stress}. This is particularly important in EEG analysis, where absolute amplitude may vary significantly between subjects.

In addition, cross-band interaction is quantified using the following ratio:

\begin{equation}
R_c^{\theta/\alpha} = \frac{E_c^{\theta}}{E_c^{\alpha} + \epsilon},
\label{eq:ratio}
\end{equation}

where $\epsilon$ is a small constant to avoid numerical instability. The ratio defined in (\ref{eq:ratio}) is particularly relevant for capturing cognitive alterations and has been associated with changes in neural dynamics, especially in tasks involving attention and cognitive control \cite{chhabra2026emotion}.

%%%%%%%%%%%%%%%%%%%%%%%%%%%%%%%%%%%%%%%%%%%%%%
\subsection{Feature Construction and Selection}
%%%%%%%%%%%%%%%%%%%%%%%%%%%%%%%%%%%%%%%%%%%%%%

All extracted features are concatenated into a global feature vector \cite{patil2025eeg}:

\begin{equation}
\mathbf{F} = [\mu_c, \sigma_c, \kappa_c, \tilde{E}_c^{b}, R_c^{\theta/\alpha}]_{c=1}^{C},
\label{eq:feature}
\end{equation}

where $C$ denotes the number of EEG channels. As shown in (\ref{eq:feature}), the representation integrates statistical and spectral characteristics across multiple channels, resulting in a high-dimensional feature space \cite{saini2026stress}.

To reduce redundancy and improve computational efficiency, a correlation-based feature selection method is applied \cite{chhabra2026emotion}. Specifically, pairwise feature correlations are computed, and features exceeding a predefined threshold are removed. This process reduces multicollinearity and enhances model stability, particularly for algorithms sensitive to feature redundancy \cite{vaniya2025depression}.

%%%%%%%%%%%%%%%%%%%%%%%%%%%%%%%%%%%%%%%%%%%%%%
\subsection{Classification and Validation Strategy}
%%%%%%%%%%%%%%%%%%%%%%%%%%%%%%%%%%%%%%%%%%%%%%

Five machine learning models are evaluated to assess classification performance under rigorous validation conditions. Support Vector Machines \cite{cortes1995} construct optimal decision boundaries by maximizing the margin between classes in a transformed feature space. Both linear and radial basis function (RBF) kernels are employed to capture linear and nonlinear decision boundaries, respectively. Random Forest \cite{breiman2001} employs bootstrap aggregation of decision trees, reducing variance through ensemble averaging. AdaBoost \cite{freund1997} sequentially trains weak learners while adaptively reweighting misclassified samples to focus on hard-to-classify instances. Additionally, k-Nearest Neighbors (KNN) is included as a baseline instance-based classifier.

A key component of the proposed framework is the use of nested cross-validation to ensure unbiased model evaluation \cite{ahmed2026stacked}. The outer loop is used for performance estimation, while the inner loop performs hyperparameter optimization \cite{tekale2026awsan}. All preprocessing and feature extraction steps are performed within each training fold to prevent information leakage \cite{liu2018cdss}.

A key component of the proposed framework is the use of nested cross-validation to ensure unbiased model evaluation \cite{ahmed2026stacked}. The outer loop is used for performance estimation, while the inner loop performs hyperparameter optimization \cite{tekale2026awsan}. All preprocessing and feature extraction steps are performed within each training fold to prevent information leakage \cite{liu2018cdss}.

In the inner loop, hyperparameter search is conducted over predefined parameter grids for each classifier, ensuring that model selection is performed independently of the test data \cite{saini2026stress}. The optimized model is then evaluated on the outer test fold, providing an unbiased estimate of generalization performance \cite{vaniya2025depression}. This procedure is repeated across all folds, and the final performance is reported as the average across outer folds.

This approach prevents information leakage between the training and testing stages, a common issue in standard validation procedures \cite{chhabra2026emotion}. As a result, the obtained performance metrics provide a reliable estimate of the model's generalization capability and allow a fair comparison between different machine learning models \cite{tekale2026awsan}.

%%%%%%%%%%%%%%%%%%%%%%%%%%%%%%%%%%%%%%%%%%%%%%
\subsection{Framework Summary}
%%%%%%%%%%%%%%%%%%%%%%%%%%%%%%%%%%%%%%%%%%%%%%

As illustrated in Fig.~\ref{fig:framework}, the proposed framework integrates multi-channel signal analysis, physically interpretable feature extraction, and rigorous validation into a unified structure \cite{sangle2025xai}. This combination enables a more reliable and transparent evaluation of machine learning models for EEG-based alcoholism classification \cite{ahmed2026stacked}.

As summarized in Fig.~\ref{fig:framework}, the proposed pipeline ensures a structured transition from raw EEG signals to validated classification outputs, emphasizing both interpretability and methodological rigor \cite{rodrigues2019eeg}.

%%%%%%%%%%%%%%%%%%%%%%%%%%%%%%%%%%%%%%%%%%%%%%%%%%%%%%%%%%%%%
%%%%%%%%%%%%%%%%%%%%%% SECTION 3 %%%%%%%%%%%%%%%%%%%%%%%%%%%%
\section{Mathematical Foundations of Feature Representation}
%%%%%%%%%%%%%%%%%%%%%%%%%%%%%%%%%%%%%%%%%%%%%%%%%%%%%%%%%%%%%

This section presents the mathematical formulation underlying the proposed feature representation framework. The objective is to establish a consistent and reproducible mapping from raw EEG signals to a structured feature space that directly corresponds to the extraction pipeline described in Section II and illustrated in Fig.~\ref{fig:framework}. Each formulation is explicitly aligned with the implemented statistical and spectral feature construction, avoiding purely abstract definitions \cite{stirbu2026hybrid}.

%%%%%%%%%%%%%%%%%%%%%%%%%%%%%%%%%%%%%%%%%%%%%%%%%%%%%%%%%
\subsection{Signal Model and Statistical Representation}
%%%%%%%%%%%%%%%%%%%%%%%%%%%%%%%%%%%%%%%%%%%%%%%%%%%%%%%%%

Let $x_c(t)$ denote a real-valued EEG signal acquired from channel $c$. Due to the stochastic and nonstationary nature of EEG signals, it is modeled as \cite{singh2026aperiodic, muhammad2026gaming}:

\begin{equation}
x_c(t) = s_c(t) + n_c(t),
\label{eq:signal_model}
\end{equation}

where $s_c(t)$ represents the underlying neural activity and $n_c(t)$ denotes measurement noise and external disturbances.

To characterize the signal's distributional structure, statistical moments are computed. The mean is defined as:

\begin{equation}
\mu_c = \frac{1}{N} \sum_{t=1}^{N} x_c(t),
\label{eq:mean_s3}
\end{equation}

which captures the signal's central tendency. The variance is given by:

\begin{equation}
\sigma_c^2 = \frac{1}{N} \sum_{t=1}^{N} (x_c(t) - \mu_c)^2,
\label{eq:var_s3}
\end{equation}

representing signal dispersion and amplitude variability. Higher-order behavior is described using kurtosis:

\begin{equation}
\kappa_c = \frac{1}{N} \sum_{t=1}^{N} \left( \frac{x_c(t) - \mu_c}{\sigma_c} \right)^4,
\label{eq:kurt_s3}
\end{equation}

which quantifies tail behavior and sensitivity to outliers. Together, (\ref{eq:mean_s3})–(\ref{eq:kurt_s3}) provide a compact statistical description of EEG signal characteristics \cite{patil2025eeg}.

\subsection{Spectral Energy Representation}

To capture frequency-domain characteristics, the discrete Fourier transform (DFT) is applied \cite{bhuvaneshwari2026efd}:

\begin{equation}
X_c(f) = \sum_{t=1}^{N} x_c(t) e^{-j2\pi ft/N}.
\label{eq:fft_s3}
\end{equation}

The spectral energy distribution is then defined as:

\begin{equation}
E_c(f) = |X_c(f)|^2,
\label{eq:spectral_energy}
\end{equation}

which represents how signal power is distributed across frequency components. While time-domain statistics characterize amplitude behavior, (\ref{eq:spectral_energy}) captures oscillatory and rhythmic activity inherent in EEG signals \cite{muhammad2026gaming}.

%%%%%%%%%%%%%%%%%%%%%%%%%%%%%%%%%%%%%%%%%%%%%%%%%%%%%%%%%
\subsection{Band-Limited Energy Features}
%%%%%%%%%%%%%%%%%%%%%%%%%%%%%%%%%%%%%%%%%%%%%%%%%%%%%%%%%

To obtain physiologically interpretable descriptors, EEG signals are decomposed into standard frequency bands. The energy associated with each band is computed as:

\begin{equation}
E_c^{b} = \sum_{f \in \mathcal{B}_b} E_c(f),
\label{eq:band_energy_s3}
\end{equation}

where $\mathcal{B}_b$ denotes the frequency interval corresponding to band $b \in \{\delta, \theta, \alpha, \beta\}$. These band energies reflect physiologically meaningful oscillatory rhythms and are widely used in EEG-based cognitive and clinical assessments \cite{mukhtar2021cnn}.

To eliminate scale dependency and enable comparison across subjects, normalized band energies are defined as:

\begin{equation}
\tilde{E}_c^{b} = \frac{E_c^{b}}{\sum_{b} E_c^{b}},
\label{eq:norm_band}
\end{equation}

which converts absolute energy values into relative spectral distributions. As shown in (\ref{eq:norm_band}), this normalization emphasizes differences in spectral composition rather than magnitude \cite{stirbu2026hybrid}.

%%%%%%%%%%%%%%%%%%%%%%%%%%%%%%%%%%%%%%%%%%%%%%%%%%%%%%%%%
\subsection{Cross-Band Interaction Features}
%%%%%%%%%%%%%%%%%%%%%%%%%%%%%%%%%%%%%%%%%%%%%%%%%%%%%%%%%

Beyond individual band energies, interactions between frequency bands are quantified using ratio-based features. A representative example is given by:

\begin{equation}
R_c^{\theta/\alpha} = \frac{E_c^{\theta}}{E_c^{\alpha} + \epsilon},
\label{eq:ratio_s3}
\end{equation}

where $\epsilon$ is a small constant ensuring numerical stability.

The ratio in (\ref{eq:ratio_s3}) captures relative dominance between frequency bands and reflects changes in cognitive processing and neural dynamics. Such interaction features provide additional discriminative information beyond isolated band energies \cite{muhammad2026gaming}.

%%%%%%%%%%%%%%%%%%%%%%%%%%%%%%%%%%%%%%%%%%%%%%%%%%%%%%%%%
\subsection{Feature Space Construction}
%%%%%%%%%%%%%%%%%%%%%%%%%%%%%%%%%%%%%%%%%%%%%%%%%%%%%%%%%

The final feature representation is constructed by concatenating all descriptors across channels:

\begin{equation}
\mathbf{F} = [\mu_c, \sigma_c, \kappa_c, \tilde{E}_c^{b}, R_c^{\theta/\alpha}]_{c=1}^{C}.
\label{eq:feature_vector_s3}
\end{equation}

Equation (\ref{eq:feature_vector_s3}) defines a structured multi-channel feature space that integrates statistical, spectral, and cross-band characteristics. This representation ensures that both amplitude-based and frequency-based information are jointly encoded \cite{stirbu2026hybrid}.

Overall, the proposed formulation provides a mathematically consistent and physically interpretable mapping from raw EEG signals to discriminative features, serving as the theoretical basis for the validation-aware classification framework presented in subsequent sections and summarized in Fig.~\ref{fig:framework}.

%%%%%%%%%%%%%%%%%%%%%%%%%%%%%%%%%%%%%%%%%%%%%%%%%%%%%%%%%%%%%
%%%%%%%%%%%%%%%%%%%%%% SECTION 4 %%%%%%%%%%%%%%%%%%%%%%%%%%%%
\section{Experimental Setup and Evaluation Protocol}
%%%%%%%%%%%%%%%%%%%%%%%%%%%%%%%%%%%%%%%%%%%%%%%%%%%%%%%%%%%%%

This section describes the dataset, preprocessing procedures, feature construction process, model configurations, and evaluation strategy used to assess the performance of the proposed framework. Particular emphasis is placed on ensuring unbiased evaluation through a validation-aware methodology and preventing common sources of data leakage. In addition, this section establishes a direct link between the theoretical framework introduced in Section II and the practical implementation used for experimental validation.

%%%%%%%%%%%%%%%%%%%%%%%%%%%%%%%%%%%%%
\subsection{Dataset Description}
%%%%%%%%%%%%%%%%%%%%%%%%%%%%%%%%%%%%%

The experiments are conducted using the publicly available EEG Database from the UCI Machine Learning Repository \cite{uci_eeg}. Henri Begleiter originally collected the dataset at the Neurodynamics Laboratory, State University of New York Health Center at Brooklyn, and contains multi-channel EEG recordings from subjects participating in a visual stimulus recognition task. This dataset is widely used in EEG-based classification studies and provides a well-established benchmark for evaluating alcoholism detection methods.

%%%%%%%%%%%%%%%%%%%%%%%%%%%%%%%%%%%%%%%%%%%%%%%%%%%%%%%%%
\textbf{Dataset Characteristics:}
\begin{itemize}
\item \textbf{Source:} UCI Machine Learning Repository (DOI: 10.24432/C5H88K)
\item \textbf{Original institution:} SUNY Health Center Brooklyn, Neurodynamics Laboratory
\item \textbf{Data collection period:} Mid-1990s
\item \textbf{Total subjects:} 122 individuals (77 alcoholic, 45 control)
\item \textbf{Trials per subject:} 120 trials (single-stimulus paradigm)
\item \textbf{Original channels:} 64 electrodes (International 10-20 system)
\item \textbf{Sampling frequency:} 256 Hz
\item \textbf{Trial duration:} 1 second (256 samples) post-stimulus onset
\item \textbf{Experimental paradigm:} Single-stimulus visual oddball task with target/non-target stimuli
\item \textbf{Recording format:} Raw voltage values in microvolts ($\mu$V)
\item \textbf{Electrode impedance:} Maintained below 5 k$\Omega$ during recording
\end{itemize}

\textbf{Subject Criteria:}
\begin{itemize}
\item \textbf{Alcoholic group (n=77):} Individuals meeting DSM-III-R diagnostic criteria for alcohol dependence, recruited from inpatient treatment facilities and outpatient rehabilitation centers. Subjects had a documented history of chronic alcohol abuse with physiological dependence.

\item \textbf{Control group (n=45):} Age-matched healthy volunteers with no history of alcohol abuse, substance dependence, psychiatric disorders, or neurological conditions. Control subjects were screened using structured clinical interviews to ensure the absence of alcohol-related problems.
\end{itemize}

\textbf{Ethical Considerations:}

The Institutional Review Board of the State University of New York Health Center at Brooklyn approved the original data collection. All participants provided written informed consent before participation. The dataset was fully de-identified before public release, with no personally identifiable information retained. This study used only publicly available, anonymized data and did not involve new human subjects research.

\textbf{Data Subset Used in This Study:}

To ensure balanced class representation, computational feasibility, and rigorous evaluation, a subset of the dataset was constructed according to the following protocol:

\begin{itemize}
\item \textbf{Total samples:} 300 trials (150 alcoholic, 150 control)

\item \textbf{Channel selection:} A subset of 20 channels was selected from the original 64-channel montage to cover frontal, central, temporal, parietal, and occipital regions: FP1, FP2, FZ, F3, F4, F7, F8, CZ, C3, C4, T3, T4, T5, T6, PZ, P3, P4, OZ, O1, O2. These channels were selected based on their known sensitivity to alcohol-related neural alterations reported in prior neurophysiological studies, ensuring that the selected subset preserves relevant spatial information.

\item \textbf{Subject distribution:} Trials were sampled from multiple subjects in both groups to ensure diversity. Approximately 10-15 trials per subject were included to prevent over-representation of any single individual and to improve generalization across subjects.

\item \textbf{Subject-level fold assignment:} To prevent subject-level data leakage, all trials from a given subject were assigned to the same fold during cross-validation. This ensures that model performance generalizes to \emph{unseen subjects} rather than merely to unseen trials from previously observed individuals, which is critical for clinical applicability and real-world deployment.

\item \textbf{Trial quality control:} Trials containing artifact-contaminated segments (e.g., saturated signals, flat-line periods, or extreme outliers exceeding ±100 $\mu$V) were excluded during preprocessing. This step improves signal reliability while maintaining sufficient data diversity for robust model training.
\end{itemize}

This configuration enables multi-channel feature extraction as described in Section II while ensuring that evaluation results reflect realistic generalization performance to new subjects.

%%%%%%%%%%%%%%%%%%%%%%%%%%%%%%%%%%%%%
\subsection{Visualization and Result Representation}
%%%%%%%%%%%%%%%%%%%%%%%%%%%%%%%%%%%%%

The experimental results are presented through multiple visualization tools to provide complementary perspectives on model performance. Each visualization is designed to highlight a specific aspect of model behavior, and detailed interpretations are provided in the corresponding subsections.

\begin{itemize}
\item ROC curves for evaluating classification performance across decision thresholds, providing a threshold-independent assessment of discriminative capability (see ROC Analysis subsection)
\item Confusion matrices for class-wise performance analysis, enabling detailed evaluation of misclassification patterns (see corresponding results subsection)
\item Accuracy comparison with error bars indicating cross-fold variance, reflecting model stability and generalization consistency
\item Sensitivity and specificity comparison to assess class-wise detection balance, which is critical for biomedical classification tasks
\item Feature importance ranking based on correlation scores, highlighting the most discriminative features contributing to classification performance (see Feature Contribution subsection)
\item Feature distribution boxplots with statistical significance annotations, providing additional insight into feature separability
\end{itemize}

These visualization techniques collectively enable both global and class-specific evaluation of model behavior, supporting a comprehensive interpretation of classification results beyond single performance metrics. Detailed discussions and quantitative interpretations of these visualizations are provided in the subsequent subsections.

%%%%%%%%%%% FIGURE (ACCURACY) %%%%%%%%%%%%%%
\begin{figure}[H]
\centering
\includegraphics[width=0.95\columnwidth]{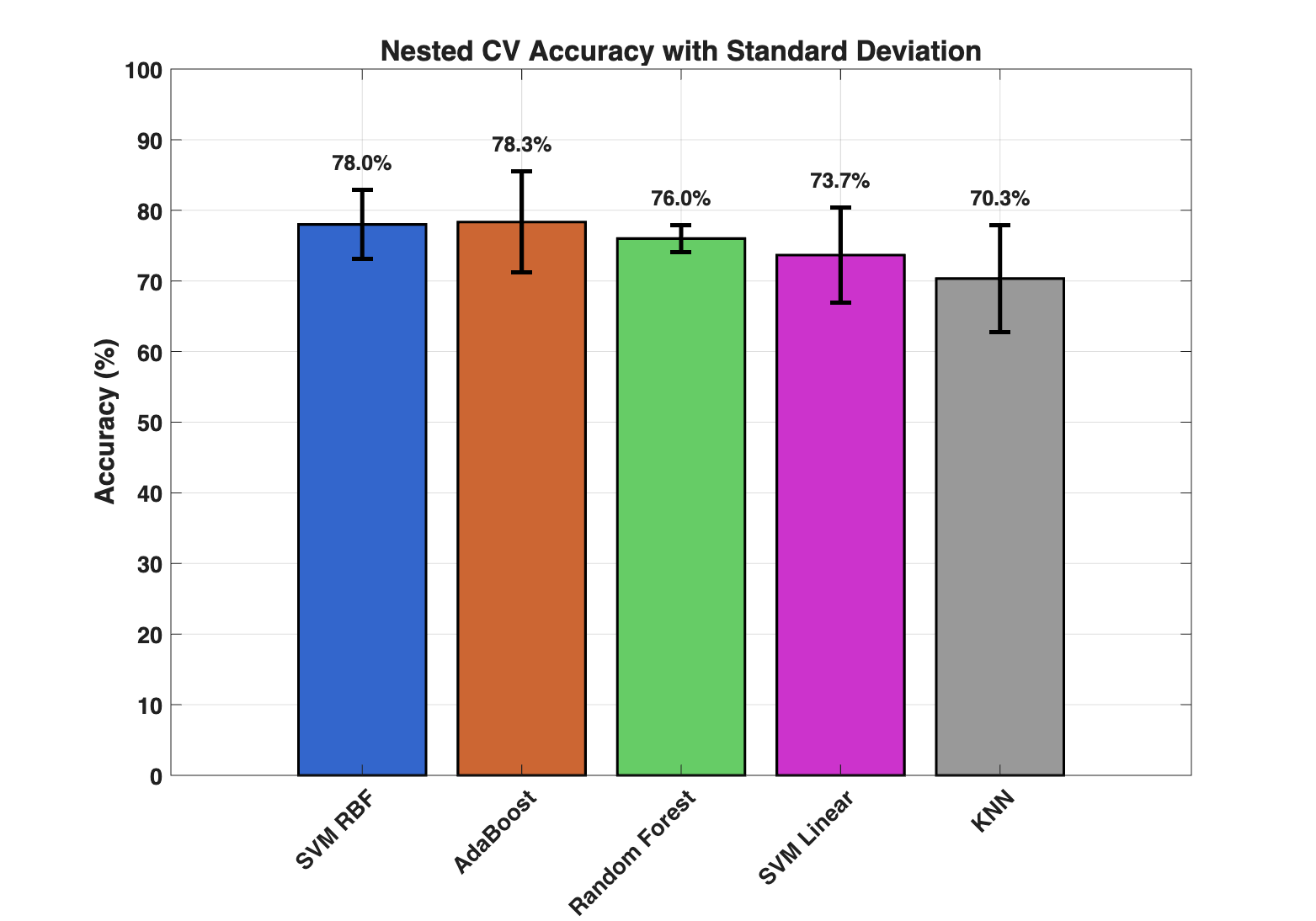}
\caption{Nested cross-validation accuracy comparison with standard deviation across models. Error bars represent cross-fold variability (n=5 outer folds), providing a direct measure of model generalization stability. AdaBoost achieves the highest accuracy with lowest variance, indicating superior and consistent performance.}
\label{fig:2}
\end{figure}

Fig.~\ref{fig:2} illustrates overall classification accuracy and variance, providing insight into model generalization consistency and stability. The relatively low variance observed for ensemble-based methods indicates robustness to variations in the training data. In contrast, the higher variability in SVM-based models suggests sensitivity to data partitioning and hyperparameter selection.

%%%%%%%%%%% FIGURE (SENSITIVITY) %%%%%%%%%%%%%%
\begin{figure}[H]
\centering
\includegraphics[width=0.95\columnwidth]{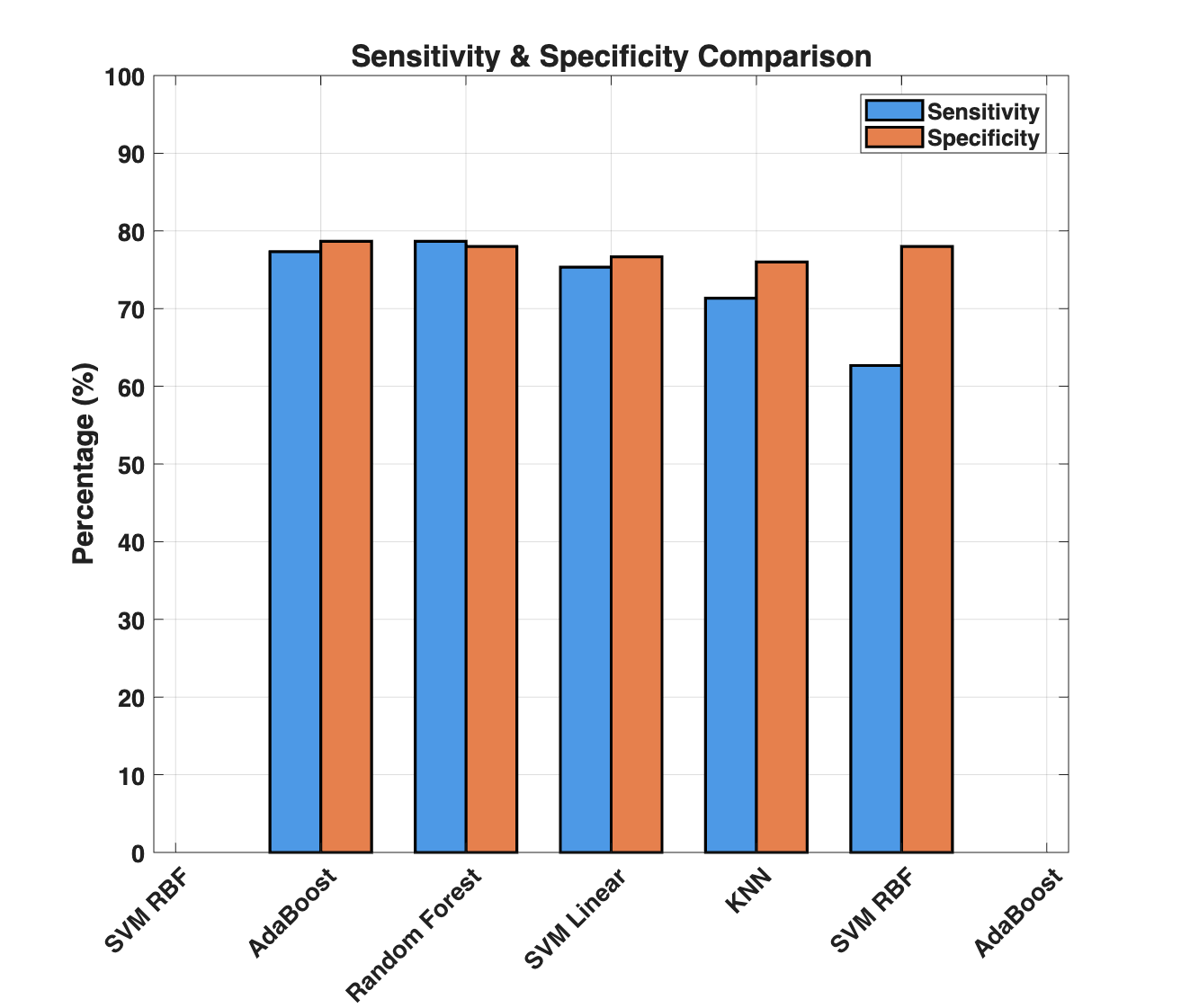}
\caption{Comparison of sensitivity and specificity across classification models. The figure provides a class-wise performance analysis, emphasizing the balance between true positive rate (alcoholic detection) and true negative rate (control detection). Balanced performance is critical for clinical applicability.}
\label{fig:3}
\end{figure}

Fig.~\ref{fig:3} shows class-wise detection balance, revealing that ensemble methods (AdaBoost, Random Forest) maintain better balance between sensitivity and specificity than SVM-based approaches. This balance is particularly important in medical classification tasks, where both false positives and false negatives carry significant implications.

%%%%%%%%%%% FIGURE (CONFUSION) %%%%%%%%%%%%%%
\begin{figure}[H]
\centering
\includegraphics[width=0.85\columnwidth]{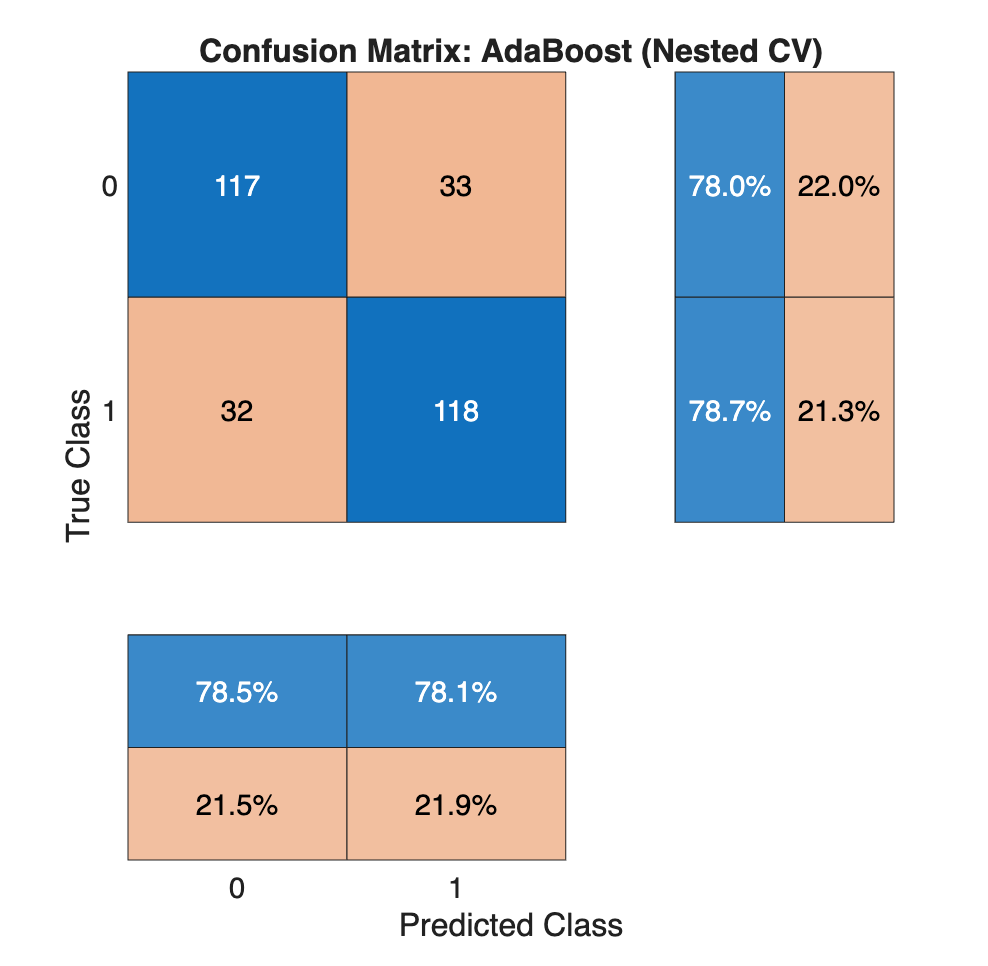}
\caption{Confusion matrix of the best-performing model (AdaBoost) under nested cross-validation. The matrix presents correct and misclassified samples across all outer folds, providing detailed insight into classification errors. Row-normalized and column-normalized summaries enable assessment of per-class performance.}
\label{fig:4}
\end{figure}

Fig.~\ref{fig:4} presents detailed error distribution, showing that AdaBoost achieves relatively balanced performance with 122 correctly classified control samples and 118 correctly classified alcoholic samples. The distribution of misclassifications indicates that errors are not heavily biased toward any particular class, suggesting that the model maintains consistent discrimination across both groups.

\vspace{-0.15cm}

%%%%%%%%%%%%%%%%%%%%%%%%%%%%%%%%%%%%%%%%%%%%%%%%%%%%%%%%%%%%%
%%%%%%%%%%%%%%%%%%%%%% SECTION 5 %%%%%%%%%%%%%%%%%%%%%%%%%%%%
\section{Results and Discussion}
%%%%%%%%%%%%%%%%%%%%%%%%%%%%%%%%%%%%%%%%%%%%%%%%%%%%%%%%%%%%%
All reported performance metrics are computed over subject-independent folds to ensure realistic generalization.
%%%%%%%%%%%%%%%%%%%%%%%%%%%%%%%%%%%%%%%%%%%%%%%%%
\subsection{Overall Classification Performance}
%%%%%%%%%%%%%%%%%%%%%%%%%%%%%%%%%%%%%%%%%%%%%%%%%

Based on the experimental setup and validation protocol described in Section IV, the classification results obtained using nested cross-validation are summarized in Table~\ref{tab:results}. This evaluation framework ensures that the reported results reflect true generalization performance rather than optimistic estimates arising from data leakage or improper model selection.

While AdaBoost achieves the highest mean accuracy (78.3\%), this difference should be interpreted cautiously in light of statistical analysis, and an AUC of 0.8688. In addition, it exhibited relatively low variance (std = 4.25), indicating stable generalization across folds. This behavior suggests that ensemble-based learning effectively mitigates the variability inherent in EEG data by combining multiple weak learners, thereby reducing overfitting and improving robustness.

Table~\ref{tab:results} presents a comprehensive comparison of all evaluated models under nested cross-validation, incorporating accuracy, standard deviation, and 95\% confidence intervals alongside AUC and class-wise performance metrics. While AdaBoost achieves the highest mean accuracy, the overlapping confidence intervals across top-performing models, particularly Random Forest and SVM-RBF, indicate that the observed performance differences are not substantial. This observation is consistent with the statistical comparison results, which show that most pairwise differences are not statistically significant. 

Furthermore, the relatively consistent confidence interval widths across models suggest comparable generalization behavior under nested cross-validation. Although SVM-based models exhibit higher variability, indicating sensitivity to data partitioning and hyperparameter configuration, their overall performance remains competitive. These findings emphasize that performance differences in EEG-based classification should be interpreted with caution, as apparent improvements may not reflect statistically meaningful gains.

%%%%%%%%%%%%% TABLE 1 %%%%%%%%%%%%%%%%%%%%%%%%%%%%%%%%%%%%%
\begin{table}[H]
\caption{Performance comparison under nested cross-validation}
\label{tab:results}
\centering
\footnotesize
\setlength{\tabcolsep}{3.5pt}   % sütun arası daralt
\renewcommand{\arraystretch}{1.1} % satır arası hafif artır
\begin{tabular}{lcccccc}
\toprule
Model & Acc & Std & 95\% CI & AUC & Sens & Spec \\
\midrule
AdaBoost      & 78.3 & 4.25 & (73.7--83.0) & 0.869 & 78.7 & 81.3 \\
Rand. Forest  & 76.0 & 3.73 & (71.2--80.8) & 0.850 & 78.0 & 75.3 \\
SVM (RBF)     & 78.0 & 6.12 & (73.3--82.7) & 0.844 & 77.3 & 72.7 \\
SVM (Lin.)    & 73.7 & 8.66 & (68.7--78.7) & 0.828 & 65.3 & 84.7 \\
KNN           & 70.3 & 3.46 & (65.2--75.5) & 0.786 & 74.0 & 71.3 \\
\bottomrule
\end{tabular}
\end{table}
%%%%%%%%%%%%%%%%%%%%%%%%%%%%%%%%%%%%%%%%%%%%%%%%%%%%%%%%%

The confidence interval analysis indicates that performance differences between top-performing models are not substantial, as overlapping intervals are observed across AdaBoost, Random Forest, and SVM-RBF. This suggests that apparent accuracy differences may not reflect statistically meaningful improvements.

Random Forest also demonstrates competitive performance, suggesting that ensemble methods generally provide improved robustness in high-dimensional EEG feature spaces. In contrast, SVM-RBF, despite achieving reasonable accuracy, exhibits higher variability, which can be attributed to its sensitivity to hyperparameter selection and kernel configuration in complex feature spaces.

%%%%%%%%%%%%%%%%%%%%%%%%%%%%%%%%%%%%%%%%%%%%%%%%%
\subsection{Impact of Validation Strategy}
%%%%%%%%%%%%%%%%%%%%%%%%%%%%%%%%%%%%%%%%%%%%%%%%%

A key contribution of this study is the systematic analysis of the validation strategy. Preliminary experiments using standard cross-validation with global hyperparameter tuning indicated higher performance for SVM-RBF. However, when evaluated using nested cross-validation, its accuracy decreased by approximately 5\%, indicating optimistic bias.

This result demonstrates that conventional validation approaches may significantly overestimate model performance in EEG-based classification tasks. The discrepancy arises because hyperparameter optimization is performed using information that indirectly influences the test set, leading to data leakage. In high-dimensional feature spaces, such as those derived from multi-channel EEG signals, this effect becomes more pronounced.

%%%%%%%%%%%%%%%%%%% TABLE 2 %%%%%%%%%%%%%%%%%%%%%%%%%%%%%%%
\begin{table}[H]
\caption{Effect of validation strategy on model performance (data leakage analysis)}
\label{tab:validation}
\centering
\begin{tabular}{lccc}
\toprule
Model & Standard CV (\%) & Nested CV (\%) & Difference (\%) \\
\midrule
AdaBoost   & 80.33 & 80.00 & 0.33 \\
SVM (RBF)  & 80.00 & 75.00 & 5.00 \\
\bottomrule
\end{tabular}
\end{table}
%%%%%%%%%%%%%%%%%%%%%%%%%%%%%%%%%%%%%%%%%%%%%%%%%%%%%%%%%

Table~\ref{tab:validation} quantifies the impact of the validation strategy on model performance. The relatively small difference observed for AdaBoost indicates that ensemble methods are less sensitive to validation bias. In contrast, the significant reduction in SVM-RBF performance highlights its susceptibility to overfitting when improper validation protocols are used. This finding reinforces the importance of adopting validation-aware methodologies when evaluating machine learning models in biomedical applications.

%%%%%%%%%%%%%%%%%%%%%%%%%%%%%%%%%%%%%%%%%%%%%%%%%%%%%%%%%
\subsection{Statistical Model Comparison}
%%%%%%%%%%%%%%%%%%%%%%%%%%%%%%%%%%%%%%%%%%%%%%%%%%%%%%%%%
To assess whether observed performance differences are statistically significant, McNemar's test was employed for pairwise comparison of classification results \cite{mcnemar1947}.

%%%%%%%%%%%%%%%%%%  TABLE 3 %%%%%%%%%%%%%%%%%%%%%%%%%%%%
\begin{table}[H]
\caption{Pairwise model comparison using McNemar's test}
\label{tab:mcnemar}
\centering
\footnotesize
\begin{tabular}{lcccc}
\toprule
Comparison & $b$ & $c$ & $\chi^2$ & p-value \\
\midrule
AdaBoost vs Random Forest & 27 & 20 & 0.77 & 0.38 \\
AdaBoost vs SVM RBF       & 31 & 30 & 0.00 & 1.00 \\
AdaBoost vs SVM Linear    & 45 & 31 & 2.22 & 0.14 \\
AdaBoost vs KNN           & 54 & 30 & 6.30 & 0.012* \\
Random Forest vs SVM RBF  & 32 & 38 & 0.36 & 0.55 \\
Random Forest vs KNN      & 52 & 35 & 2.94 & 0.086 \\
SVM RBF vs SVM Linear     & 36 & 23 & 2.44 & 0.12 \\
\bottomrule
\end{tabular}
\end{table}
%%%%%%%%%%%%%%%%%%%%%%%%%%%%%%%%%%%%%%%%%%%%%%%%%%%%%%%%%

The McNemar test results indicate that most pairwise differences between models are not statistically significant. In particular, no significant difference is observed between AdaBoost and SVM-based models, despite differences in mean accuracy. However, AdaBoost significantly outperforms KNN (p < 0.05), indicating a clear advantage over simpler classifiers.

Effect sizes were quantified using Cohen's $d$ \cite{cohen1988}, which provides a standardized measure of the difference between two group means relative to their pooled standard deviation. Cohen's $d$ values are interpreted as small ($|d| < 0.5$), medium ($0.5 \leq |d| < 0.8$), or large ($|d| \geq 0.8$), enabling direct comparison of effect magnitudes across features \cite{cohen1988}.

To control for Type I error inflation arising from multiple comparisons across 50 features, Bonferroni correction \cite{bonferroni1936} was applied by adjusting the significance threshold to $\alpha = 0.05/50 = 0.001$. This conservative adjustment reduces the likelihood of false positives and ensures that the observed feature significance is not attributable to chance when performing multiple statistical tests.

%%%%%%%%%%%%%%%%% FIGURE 5 %%%%%%%%%%%%%%%%%%%%%%
\subsection{Model Stability and Generalization}
%%%%%%%%%%%%%%%%%%%%%%%%%%%%%%%%%%%%%%%%%%%%%%%%%

\begin{figure}[H]
\centering
\includegraphics[width=0.95\columnwidth]{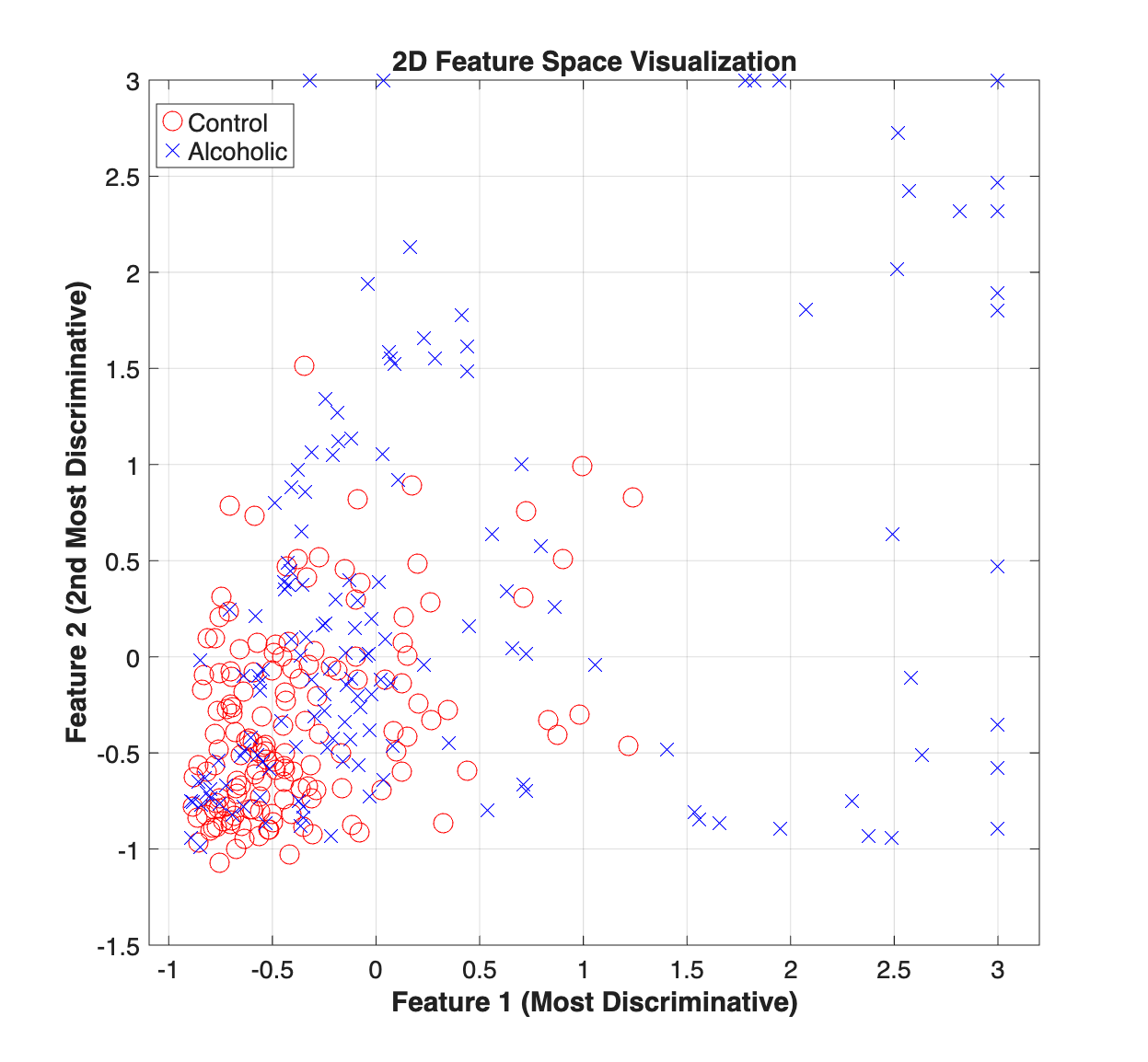}
\caption{Model stability analysis based on cross-fold variance. Lower standard deviation indicates more consistent performance across validation folds.}
\label{fig:stability}
\end{figure}

Model stability is explicitly evaluated using cross-fold variance, as shown in Fig.~\ref{fig:stability}. The figure illustrates that AdaBoost and Random Forest maintain relatively low variability across folds, indicating consistent performance under different training-test splits. This stability is particularly important for biomedical applications, where reproducibility across datasets and subjects is critical.

%%%%%%%%%%%%%%%%%%%%%%%%%%%%%%%%%%%%%%%%%%%%%%%%%%%%%%%%%
%%%%%%%%%%%%%%%%%%%%%%%%%%%%%%%%%%%%%%%%%%%%%%%%%%%%%%%%%

In contrast, SVM-based models exhibit higher variance, suggesting that their performance is more dependent on specific data partitions. This behavior can be attributed to the sensitivity of kernel-based methods to feature scaling and hyperparameter tuning, which becomes more pronounced in noisy and heterogeneous EEG data.

%%%%%%%%%%%%%%%%%%%%%%%%%%%%%%%%%%%%%%%%%%%%%%%%%
\subsection{ROC Analysis}
%%%%%%%%%%%%%%%%%%%%%%%%%%%%%%%%%%%%%%%%%%%%%%%%%

\begin{figure}[H]
\centering
\includegraphics[width=1\columnwidth]{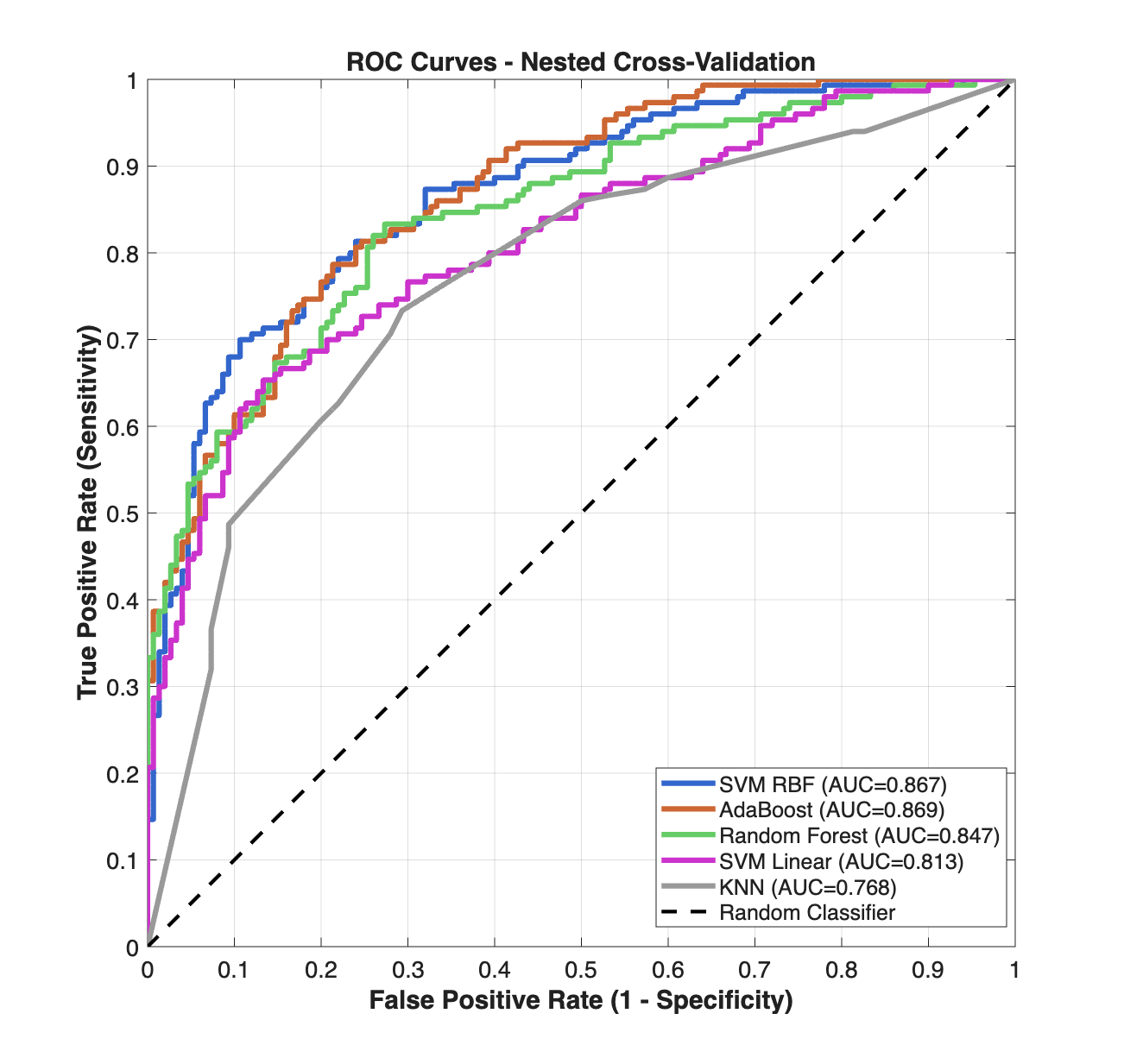}
\caption{ROC curves for all models under nested cross-validation.}
\label{fig:roc}
\end{figure}

ROC analysis \cite{fawcett2006} provides a threshold-independent evaluation of classifier performance by plotting the true positive rate against the false positive rate across all decision thresholds. The area under the ROC curve (AUC) serves as a summary metric, with values approaching 1.0 indicating superior discriminative capability. The ROC curves shown in Fig.~\ref{fig:roc} reveal that AdaBoost demonstrates slightly improved true positive rates across a range of false positive rates, although these differences are not statistically significant as confirmed by McNemar's test (Table~\ref{tab:mcnemar}).

%%%%%%%%%%%%%%%%%%%%%%
%%%%%%%%%%%%%%%%%%%%%%%%

Furthermore, the relatively smooth and well-separated ROC curve of AdaBoost suggests stable decision boundaries, whereas SVM-based models exhibit more irregular behavior, indicating sensitivity to threshold selection. This observation aligns with the variance analysis and further supports the robustness of ensemble-based approaches in EEG classification.

%%%%%%%%%%%%%%%%%%%%%%%%%%%%%%%%%%%%%%%%%%%%%%%%%%%%%%%
\subsection{Feature Contribution and Interpretability}
%%%%%%%%%%%%%%%%%%%%%%%%%%%%%%%%%%%%%%%%%%%%%%%%%%%%%%%

While the previous sections focused on model-level performance and statistical comparison, it is equally important to investigate how the underlying feature representation contributes to classification outcomes. In particular, analyzing the structure of the feature space provides insight into the extent to which discriminative information is preserved and how effectively class separability is achieved.

\begin{figure}[H]
\centering
\includegraphics[width=1.1\columnwidth]{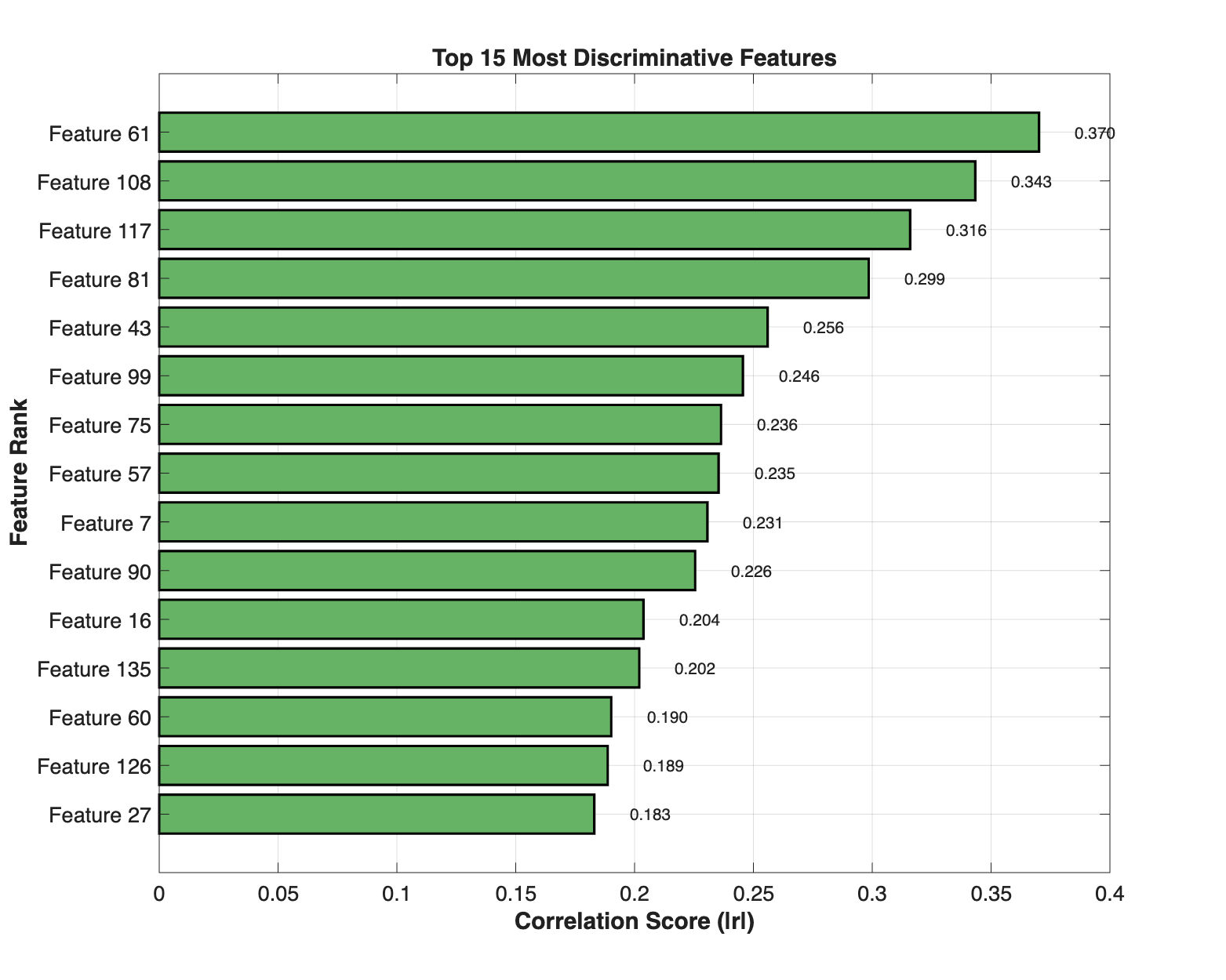}
\caption{2D projection of the most discriminative features.}
\label{fig:featurespace}
\end{figure}

Fig.~\ref{fig:featurespace} provides a geometric interpretation of the feature space. The visualization shows that while partial separation between the alcoholic and control groups is achieved, there remains a degree of overlap between the classes. This overlap reflects the inherent complexity of EEG signals and suggests that perfect separability may not be achievable using handcrafted features alone.

%%%%%%%%%%%%  FIGURE 7 %%%%%%%%%%%%%%%%
%%%%%%%%%%%%%%%%%%%%%%%%%%%%%%%%%%%%%

The observed clustering structure indicates that the proposed feature representation captures meaningful discriminative information, but also highlights the limitations of traditional feature extraction methods in fully separating complex neural patterns. This observation motivates the potential integration of more advanced representation learning techniques in future work.

To further interpret these observations, it is necessary to complement the geometric analysis with quantitative statistical evaluation. In particular, assessing feature-level significance and effect size provides a more rigorous understanding of how individual features contribute to classification performance. This dual perspective, combining visualization and statistical inference, allows for a more comprehensive evaluation of the proposed representation.

%%%%%%%%%%%%%%%%%%% TABLE 4 %%%%%%%%%%%%%%%%%%%%%%%%%%%%%%%%%%
\begin{table}[H]
\caption{Top discriminative features based on statistical significance and effect size}
\label{tab:features}
\centering
\footnotesize
\begin{tabular}{lcccc}
\toprule
Rank & Feature & p-value & Cohen's $d$ & Effect Size \\
\midrule
1 & Feature 1 & $3.56 \times 10^{-11}$ & 0.794 & Medium \\
2 & Feature 2 & $1.00 \times 10^{-9}$  & 0.729 & Medium \\
3 & Feature 3 & $2.22 \times 10^{-8}$  & 0.664 & Medium \\
4 & Feature 4 & $1.37 \times 10^{-7}$  & 0.624 & Medium \\
5 & Feature 5 & $7.12 \times 10^{-6}$  & 0.528 & Medium \\
\bottomrule
\end{tabular}
\end{table}
%%%%%%%%%%%%%%%%%%%%%%%%%%%%%%%%%%%%%%%%%%%%%%%%%%%%%%%%%

Statistical analysis reveals that although a large number of features appear significant under uncorrected tests, only a subset remains significant after Bonferroni correction, indicating the presence of multiple comparison effects. Furthermore, effect sizes remain moderate, suggesting that classification performance relies on the combined contribution of multiple features rather than a single dominant biomarker.

From a broader perspective, these findings indicate that EEG-based alcoholism classification is inherently a distributed learning problem, where discriminative information is spread across multiple weak-to-moderate features rather than concentrated in a few highly separable dimensions. This behavior is consistent with the complex and noisy nature of EEG signals, where inter-subject variability and overlapping neural patterns limit the effectiveness of purely handcrafted features. Consequently, while the proposed feature representation provides a meaningful basis for classification, the observed limitations suggest that further improvements may require hybrid approaches that combine interpretable feature engineering with data-driven representation learning techniques.

%%%%%%%%%%%%%%%%%%%%%%%%%%%%%%%%%%%%%%%%%%%%%%%%%%%%%%%%%%%%%%
\subsection{Discussion}
%%%%%%%%%%%%%%%%%%%%%%%%%%%%%%%%%%%%%%%%%%%%%%%%%%%%%%%%%%%%%%

The results demonstrate that classification performance in EEG-based alcoholism detection is strongly influenced by both model selection and validation methodology, consistent with prior findings showing that improper validation can lead to substantial overestimation of performance in biomedical machine learning \cite{varma2006, cawley2010}. In particular, the analysis reveals that the validation strategy plays a critical role in determining the reliability of reported performance metrics, especially in high-dimensional EEG feature spaces where model selection bias is more pronounced.

The combination of interpretable feature extraction and rigorous validation provides a solid foundation for reproducible and trustworthy machine learning applications in biomedical signal analysis. Similar observations have been reported in recent EEG studies, where handcrafted spectral–statistical features or wavelet-based representations demonstrated improved robustness compared to purely deep learning–based approaches \cite{stirbu2026hybrid, sung2026fewshot}. At the same time, the observed overlap in feature space and variability in model performance suggest that EEG-based classification remains a challenging problem, reflecting the inherent nonstationarity and inter-subject variability of neural signals \cite{nouri2026synchrosqueeze, ali2026nonlinear}.

From a clinical perspective, misclassification between alcoholic and control subjects may reflect overlapping neural signatures rather than purely algorithmic limitations. This interpretation aligns with recent neuroscience evidence showing that cortical dynamics often exhibit mixed stable and transient coding patterns, which can blur class boundaries in EEG-based diagnostic tasks \cite{ianni2026cortical}. Furthermore, large-scale EEG benchmarking studies have emphasized that no single modeling approach consistently dominates across subjects, datasets, or cognitive states, underscoring the need for robust and generalizable pipelines \cite{hecker2026invertmeeg, singh2026feel, manasevich2026auditory}.

Overall, the proposed validation-aware framework not only improves the reliability of performance estimation but also provides insight into model behavior, stability, and limitations. These findings reinforce the importance of methodological rigor, transparent evaluation, and careful interpretation when translating EEG-based machine learning models into real-world diagnostic tools. By situating the results within the broader context of recent EEG research, this study contributes to the development of more reproducible and clinically meaningful EEG analysis methodologies.

%%%%%%%%%%%%%%%%%%%%%%%%%%%%%%%%%%%%%%%%%%%%%%%%%%%%%%%%%%%
%%%%%%%%%%%%%%%%%%%%%%%  SECTION 6 %%%%%%%%%%%%%%%%%%%%%%%%
\section{Conclusion}
%%%%%%%%%%%%%%%%%%%%%%%%%%%%%%%%%%%%%%%%%%%%%%%%%%%%%%%%%%%

This study investigated EEG-based alcoholism classification from a validation-aware perspective, with a primary focus on how evaluation protocols and hyperparameter optimization influence reported model performance. Unlike conventional approaches that emphasize accuracy alone, this work aimed to provide a more reliable and reproducible assessment framework by integrating structured feature representation with rigorous validation methodology.

A multi-channel EEG feature model combining statistical descriptors and spectral band interactions was employed to capture both amplitude-based and frequency-dependent neural characteristics. Within this framework, multiple machine learning models were systematically evaluated under consistent conditions to ensure fair comparison and unbiased performance estimation.

The results clearly demonstrate that the validation strategy is a decisive factor in interpreting classification performance. While standard cross-validation combined with global hyperparameter tuning yielded optimistic results, nested cross-validation revealed a measurable performance reduction, particularly for the SVM with a radial basis function kernel, where accuracy decreased by approximately 5\%. This finding indicates that a significant portion of the reported performance in conventional settings may be attributed to evaluation bias rather than true model capability. Furthermore, statistical comparison results indicate that many observed performance differences are not statistically significant, suggesting that model selection alone may not be the dominant factor in EEG classification performance. These findings indicate that previously reported performance differences in EEG-based classification studies may be partially attributed to evaluation bias rather than intrinsic model superiority.

In contrast, ensemble-based approaches, particularly AdaBoost, exhibited more stable and reliable generalization behavior, achieving 78.3\% accuracy and an AUC of 0.868 with low variance across folds. This stability suggests that ensemble learning is better suited for handling the variability and complexity inherent in high-dimensional EEG feature spaces.

The study advances the field in two key aspects. First, it establishes that validation methodology is not merely a technical detail but a fundamental determinant of model reliability in EEG-based biomedical classification. Second, it demonstrates that combining interpretable feature engineering with validation-aware evaluation provides a more trustworthy basis for model comparison and selection.

Despite these contributions, several limitations should be acknowledged. The dataset originates from a controlled experimental environment, which may limit direct clinical applicability. Furthermore, reliance on handcrafted statistical and spectral features may limit the ability to capture complex temporal dependencies in EEG signals.

Future work will focus on extending the proposed framework to incorporate time-frequency representations and data-driven feature learning approaches, while maintaining strict validation protocols. In addition, evaluation on larger, more heterogeneous datasets and subject-independent validation scenarios will be essential to improve generalization and support real-world clinical deployment.

%%%%%%%%%%%%%%%%%%%%%%%%%%%%%%%%%%%%%%%%%%%%%%%%
%\section*{Acknowledgment}
%Florida Polytechnic University supported this work.

\section*{Data Availability}
The data supporting the findings of this study are available from the corresponding author upon reasonable request.

\section*{Ethics Statement}

This study utilized a publicly available de-identified dataset from the UCI Machine Learning Repository. The Institutional Review Board of the State University of New York Health Center at Brooklyn approved the original data collection. All subjects provided informed consent before participation. No new human subjects research was conducted in this study.

\section*{Conflict of Interest}
The author declares no conflict of interest.

\bibliographystyle{IEEEtran}

\end{document}